\documentclass[aps,pra,twocolumn,10pt,nofootinbib,showpacs]{revtex4-1}
\usepackage{amsmath}
\usepackage{graphicx}
\usepackage{amsfonts}
\usepackage{braket}
\usepackage{natbib}
\usepackage[utf8]{inputenc}
\usepackage[colorlinks=true,citecolor=blue,linkcolor=blue]{hyperref}
\usepackage{dsfont}

\usepackage{xcolor}

\newcommand{\LR}[1]{\left(#1\right)}

\begin{document}

\title{Optimal Control of Quantum Measurement}
\author{D. J. Egger}
\author{F. K. Wilhelm}
\affiliation{Theoretical Physics, Universität des Saarlandes, D-66123 Saarbrücken, Germany}

\begin{abstract}
Pulses to steer the time evolution of quantum systems can be designed with optimal control theory. In most cases it is the coherent processes that can be controlled and one optimizes the time evolution towards a target unitary process, sometimes also in the presence of non-controllable incoherent processes. Here we show how to extend the GRAPE algorithm in the case where the incoherent processes are controllable and the target time evolution is a non-unitary quantum channel. We perform a gradient search on a fidelity measure based on Choi matrices. We illustrate our algorithm by optimizing a phase qubit measurement pulse. We show how this technique can lead to large measurement contrast close to $99\%$. We also show, within the validity of our model, that this algorithm can produce short $1.4$ ns pulses with $98.2\%$ contrast.
\end{abstract}



\maketitle

\section{Introduction}
Quantum optimal control theory is the science of shaping control pulses to manipulate quantum systems in a useful way \cite{Rice_Book, Brumer_Book}. In many quantum control systems, the time evolution is optimized under unitary time evolution. Examples of this include the evolution of many electron systems under Hamiltonian dynamics \cite{Castro_PRL_109_153603} as well as time evolution under a non-linear Schr\"odinger equation \cite{Sklarz_PRA_66_053619} or also quantum gates for quantum computing in solid state systems \cite{Egger_SUST_27_014001, Cerfontaine_arXiv, Schutjens_PRA_88_052330, Vesterinen_arXiv}. Additionally, optimization towards a target unitary time evolution can also be done in the presence of non-unitary dynamics \cite{Floether_NJP_14_073023, Herbruggen_JPB_44_154013}. However, some desired quantum processes are inherently incoherent, such as cooling \cite{Reich_NJP_15_125028}. A central application of incoherent processes is measurement within the field of circuit QED. Unlike many other detection processes in quantum physics, object and detector are made out of the same technology and act on similar time scales making careful design possible and necessary. Similar statements can be made about readout of quantum states in semiconductor quantum dots \cite{Elzerman_Nature_430_431, Gilad_PRL_97_116806, Elzerman_PRB_67_161308}. The read out mechanism depends on the type of superconducting qubit \cite{Clark_Nature_453_1031, Devoret_Science_339_1169} being used. For instance transmon qubits are typically read out through a resonator \cite{Koch_PRA_76_042319, Jeffrey_PRL_112_190504} whilst phase qubit readout is based on tunneling out of a metastable well \cite{Neeley_NatPhys_4_523, Chen_APL_101_182601}. Additionally, this tunneling mechanism can also be used to create a microwave photon counter named the Josephson photomultiplier (JPM) \cite{Chen_PRL_107_217401}. It is usually desirable to have a high measurement contrast and in some cases high speed. The latter is particularly crucial for quantum computing which can involve many measurements \cite{Fowler_PRA_86_032324}.

In this paper we expand the gradient ascent pulse engineering (GRAPE) optimal control algorithm to the optimization of non-unitary quantum channels using Choi matrices. The algorithm is presented in section \ref{Sec:algo}. We illustrate it in section \ref{Sec:PQ} with the optimization of a readout pulse for the phase qubit. Conclusions are drawn in section \ref{sec:conclusion}.

\section{Optimal Control Algorithm \label{Sec:algo}}
An open quantum system with Markovian dynamics follows the time evolution given by a Lindblad master equation \cite{Nielsen_and_Chuang_book}. The time evolved density matrix can be found by vectorizing the master equation using the identity ${\rm col}(ABC)=(C^T\otimes A){\rm col}(B)$. Here ${\rm col}(X)=\vec{X}$ denotes column stacking of the matrix $X$. The result is a first order differential equation $\dot{\vec{\rho}}=\mathcal{S}(t)\,\vec{\rho}$ for the vectorized density matrix $\vec{\rho}$ \cite{Herbruggen_JPB_44_154013}. This equation is similar to the Schr\"odinger equation and can be solved by exponentiating the generator $\mathcal{S}(t)$. The time evolution, of duration $T$, of a general initial density matrix is thus given by $\vec{\rho}(T)=\mathcal{T}(T)\vec{\rho}(0)$ with the time propagator $\mathcal{T}$ given by the time ordered exponential of the integral of the generator. For a column stacked vectorized master equation the generator is
\begin{align} \label{Eqn:PII_Chap5_Gen}
 \mathcal{S}(t)=&~i\LR{\hat H^T\otimes\mathds{1}-\mathds{1}\otimes\hat H}\\ \notag &+\sum\limits_l\gamma_l\LR{\hat L_l^*\otimes\hat L_l-\frac{1}{2}\hat L_l^T\hat L_l^*\otimes\mathds{1}-\frac{1}{2}\mathds{1}\otimes\hat L_l^\dagger\hat L_l^{\phantom{\dagger}}}
\end{align}
where $\hat H$ is the Hamiltonian and $\hat L_l$ is the Lindblad operator associated to the incoherent process with rate $\gamma_l$. Note that having the rates be positive for all times ensures that the resulting dynamics is completely positive and trace preserving \cite{Breuer_JPBAMOP_45_154001}. Within this generator are hidden the control fields $\boldsymbol u(t)$. They can be located in the Hamiltonian $\hat H$ which, as in the GRAPE algorithm \cite{Khaneja_JMR_172_296305}, is separated into drift $\hat H_\text{d}$ and controls $\hat H_k$. However they can also control some of the rates such that the set of rates can be split into controllable rates and drift rates $\{\gamma_l\}=\{\gamma_{l,\text{d}},\gamma_{l,\text{c}}(\boldsymbol u(t))\}$. This suggests a drift-control decomposition for the generator
\begin{align} \notag
 \mathcal{S}(t)=\mathcal{S}_\text{d}+\sum\limits_k f_k(\boldsymbol u(t))\mathcal{S}_k\,.
\end{align}
The drift term $\mathcal{S}_\text{d}$ is the part of Eq. (\ref{Eqn:PII_Chap5_Gen}) containing the drift Hamiltonian $\hat H_\text{d}$ and the Lindblad operators corresponding to the drift rates $\gamma_{l,\text{d}}$. The control part is the remainder of Eq. (\ref{Eqn:PII_Chap5_Gen}). It contains terms dependent on $\boldsymbol u(t)$. The functions $f_k$ account for possible non linear behaviors with respect to $\boldsymbol u(t)$. However, these functions $f_k$ are known and assumed to be differentiable allowing us to use the chain rule when computing gradients with respect to the controls. When dealing with actual experiments, fine tunning of the control pulses can be done with adaptive hybrid optimal control (Ad-HOC) if these functions are not properly characterized \cite{Egger_PRL_112_240503}.

Similarly to the GRAPE algorithm, the controls are discretized in time into $N$ piecewise constant control pixels of duration $\Delta T$ and the time propagator $\mathcal{T}(T)$ is approximated by
\begin{align} \notag
 \mathcal{T}(T)=\prod\limits_{j=N-1}^0e^{\mathcal{S}(j\Delta T)\Delta T}\,.
\end{align}
Note that in the product early times go to the right to satisfy time ordering and the product counts down. $\mathcal{S}(j\Delta T)$ is the generator evaluated at pixels $\boldsymbol u(j\Delta T)$. This time evolution corresponds to a quantum channel which we wish to optimize. To do so a fidelity measure based on Choi matrices \cite{Choi_LinAlg_10_285290, Stromer_Springer_2013} is constructed. The Choi matrix $C$ is related to the time propagator $\mathcal{T}$ by reorganizing the elements according to
\begin{align} \label{Eqn:PII_Choi2T}
 C_{d\alpha+\beta,d\alpha'+\beta'}=\mathcal{T}_{d\beta'+\beta,d\alpha'+\alpha}\,,
\end{align}
where $d$ is the dimension of the Hilbert space and $\alpha,\alpha',\beta,\beta'\in\{1,...,d\}$. This can be shown by noticing that the vectorized matrix $\ket{i}\!\!\bra{j}$ is the unit vector $\hat e_{dj+i}$ with 1 on entry $dj+i$ and zero elsewhere. Therefore with $[\mathcal{E}(\ket{i}\!\!\bra{j})]_{\beta,\beta'}=\mathcal{T}_{d\beta'+\beta,dj+i}$ and $C=\sum_{ij}\ket{i}\!\!\bra{j}\otimes\mathcal{E}(\ket{i}\!\!\bra{j})$ which defines the Choi matrix, the above identity ensues. A natural way to measure how close  the realized quantum channel is to a target channel, described by a Choi matrix $C_\text{t}$, is through the channel fidelity \cite{raginsky_PLA01_290_1118}
\begin{align} \notag
 \Phi_\text{ch}=\frac{1}{d^2}\LR{{\rm Tr}\left\{\sqrt{\sqrt{C_\text{t}}C[\boldsymbol u]\sqrt{C_\text{t}}}\right\}}^2\,.
\end{align}
This fidelity was constructed from the fidelity between two states $\rho$ and $\sigma$ given by $\mathcal{F}={\rm Tr}\sqrt{\sqrt{\rho}\sigma\sqrt{\rho}}$ \cite{Nielsen_and_Chuang_book} by using the Choi-Jamiolkwoski isomorphism which, loosely speaking, relates quantum channels to states in a higher dimension. The channel fidelity $\Phi_\text{ch}$ reduces to, in the case when both processes are unitary, to the gate overlap fidelity $\Phi_\text{\tiny QPT}=|{\rm Tr}\{\hat U_\text{t}^\dagger\hat U[\boldsymbol u]\}|^2/d^2$ where $\hat U_\text{t}$ is the target unitary matrix. However it is not suitable for a pulse optimisation algorithm due to the square root which prevents an analytical expression for the gradient. Instead we define a fidelity starting from the square of the Frobenius norm
\begin{align} \notag
 \|C_\text{t}-C[\boldsymbol u]\|^2={\rm Tr}\left\{C_\text{t}^2\right\} +&~ {\rm Tr}\left\{C[\boldsymbol u]^2\right\} \\ \notag
 &-2\,{\rm Re\,Tr}\left\{C_\text{t}^\dagger C[\boldsymbol u]\right\}\,.
\end{align}
The equality follows from the definition of the Frobenius norm. As the realized channel approaches the target one, the error $\|C_\text{t}-C[\boldsymbol u]\|^2$ is reduced. This prompts the following definition for the fidelity
\begin{align} \label{Eqn:PII_FidChannel}
 \Phi_\text{ch}'=\frac{{\rm Re\,Tr}\left\{C_\text{t}^\dagger C[\boldsymbol u]\right\}}{{\rm Re\,Tr}\left\{C_\text{t}^\dagger C_\text{t}\right\}}\,.
\end{align}
The factor in the denominator has been included to upper bound the fidelity by one. Its presence is called for by the fact that, contrary to density matrices, Choi matrices do not have unit trace. Note that this expression is not sensitive to global phases contrary to its counterpart for unitary matrices \cite{Khaneja_JMR_172_296305}. The gradient with respect to the control pixels is
\begin{align} \label{Eqn:PII_GradFidChannel1}
 \nabla_{kj}\Phi_\text{ch}'=\frac{{\rm Re\,Tr}\left\{C_\text{t}^\dagger\, \frac{\partial C[\boldsymbol u]}{\partial u_{kj}}\right\}}{{\rm Re\,Tr}\left\{C_\text{t}^\dagger C_\text{t}\right\}}\,.
\end{align}
The gradient of the Choi matrix is found by computing the gradient of the time propagator and rearranging the terms according to Eq. (\ref{Eqn:PII_Choi2T}). The procedure to compute the gradient of $\mathcal{T}$ follows the same idea as for the unitary case. However, since in a generic open system $\mathcal{S}$ is not necessarily normal \cite{Machnes_PRA_84_022305}, the procedure of computing the gradient of a single pixel using eigenvalues does not work. Instead the identity
\begin{align} \label{Eqn:PII_GradFidChannel2}
 \left.\frac{{\rm d}}{{\rm d} x}e^{A+xB}\right\vert_{x=0}=e^A\int_0^1 e^{-A\tau}Be^{A\tau}{\rm d}\tau
\end{align}
is used. The latter can be evaluated exactly using augmented matrix exponentials \cite{Floether_NJP_14_073023}
\begin{align} \label{Eqn:PII_GradFidChannel3}
 \exp\begin{pmatrix} A & B \\ 0 & A \end{pmatrix}=\begin{pmatrix} e^A & \int_0^1e^{A(1-\tau)}Be^{A\tau}{\rm d}\tau \\ 0 & e^A \end{pmatrix}\,.
\end{align}
Thus for computing $\partial \mathcal{T}/\partial u_{kj}$ one sets $A=\mathcal{S}(j\Delta T)\Delta T$ and $B=\mathcal{S}_k\Delta T$. Given that the augmented matrix can be defective, its exponential is computed with Ward's Pad\'e approximation \cite{Ward_JNA_14_600,Moler_SIAM_45_3}. Finally all elements are in place to successfully optimize the pulse of a non-unitary process towards a target non-unitary channel using the GRAPE and BFGS algorithms \cite{Fouquieres_JMP_212_412, Nocedal_Springer_2006}. The fidelity is given by Eq. (\ref{Eqn:PII_FidChannel}) whilst its gradient is found from Eqs. (\ref{Eqn:PII_GradFidChannel1}) through (\ref{Eqn:PII_GradFidChannel3}).

\section{Optimization of a Phase Qubit Measurement Pulse \label{Sec:PQ}}
The flux biased phase qubit is a superconducting circuit made of a large area Josephson junction (JJ) shunted by an inductor. Threading an external flux through this loop makes the energy levels tunable and also allows for easy readout \cite{Neeley_NatPhys_4_523, Chen_APL_101_182601}. This type of qubit can be biased in a regime where the potential is made of a shallow and a deep well. The qubit logical $\ket{0}$ and $\ket{1}$ basis is formed in the shallow well. When the qubit is read out, a flux pulse makes the shallow well shallower; the $\ket{1}$ state tunnels into the deeper well whilst tunneling of $\ket{0}$ is exponentially smaller. A tunneling event creates a flux change that can be picked-up by a nearby SQUID \cite{Cooper_PRL_93_180401}. JPMs allow single photon detection in the microwave regime and are also based on a phase qubit like device \cite{Chen_PRL_107_217401, Govia_PRA_86_032311}. Here we will show how to optimize a measurement pulse for a phase qubit using the methods described in the previous section.

\subsection{Phase Qubit Model}
The phase qubit \cite{Cooper_PRL_93_180401, Simmonds_PRL_93_077003}, flux biased by $\varphi_\text{b}$ but without current bias, is described by the Hamiltonian
\begin{align} \label{Eqn:PII_PhaseQubitH}
 \hat H=E_c\hat N^2+E_J\LR{\frac{1}{2\beta}(\hat \varphi-\varphi_\text{b})^2-\cos\hat\varphi}\,.
\end{align}
The charging energy is $E_c=2e^2/C$ and the Josephson coupling energy is $E_J=I_0\hbar/2e$. The qubit is coupled to the external bias flux $\Phi_0\varphi_\text{b}$ by the constant $\beta=2eLI_0/\hbar$. The critical current of the JJ is $I_0$ and its associated capacitance is $C$ whilst the shunt inductance is $L$.

\subsubsection{The Three Level Model}
\begin{figure} \centering
 \includegraphics[width=0.85\columnwidth]{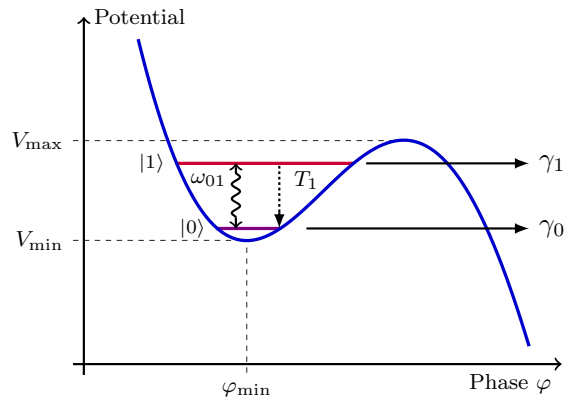}
 \caption{Sketch of the phase qubit's potential focusing on the shallow well. The wavy line indicates the $0\leftrightarrow1$ transition frequency which is a coherent process and enters in the Hamiltonian. Controllable incoherent processes are indicated by solid straight lines whereas the uncontrollable $T_1$ relaxation process is constant.  \label{Fig:PII_PhaseQubitPot}}
\end{figure}

When biased a little below $\varphi_\text{b}=2\pi$ the potential has a shallow and a deep well. The qubit states $\ket{0}$ and $\ket{1}$ are formed out of the two lowest states of the shallow well. By raising the bias closer to $2\pi$, the shallow well becomes shallower allowing the $\ket{1}$ and $\ket{0}$ states to tunnel into the deeper well, see Fig. \ref{Fig:PII_PhaseQubitPot}. Furthermore, at these bias values, the deep well is much deeper than the shallow well. Thus, the potential can be approximated by a cubic function where the deep well is treated as a continuum. This prompts a three state description of the qubit formed by the basis $\{\ket{0},\,\ket{1},\,\ket{\text{m}}\}$. $\ket{\text{m}}$ is a combination of all the states that $\ket{0}$ and $\ket{1}$ can incoherently tunnel into.

\begin{figure} \centering
 \includegraphics[width=0.95\columnwidth]{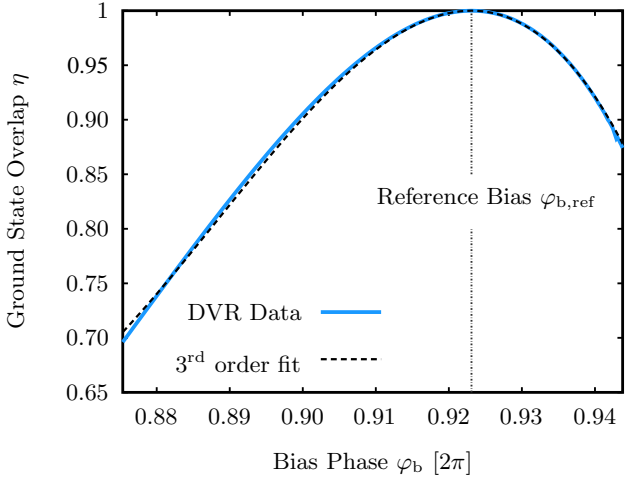}
 \caption{ $\ket{0}\leftrightarrow\ket{1}$ state mixing parameter $\eta$ as function of the bias phase. The solid line indicates numerical DVR data whilst the dashed line is a third order polynomial fit to preserve analyticity when performing the gradient search. \label{Fig:PII_3lvl_eta}}
\end{figure}

The bias flux changes the shape of the potential, thus for different $\varphi_\text{b}$ the logical $\ket{0}$ and $\ket{1}$ states have different wave functions. For an arbitrary bias flux the three level model Hamiltonian is expressed with respect to a reference bias $\varphi_\text{ref}$
\begin{align} \notag
 \hat H=P\hat H_\text{ref}P^{-1}~~~\text{with}~~~P=\begin{pmatrix} \eta & \sqrt{1-\eta^2} & 0 \\ \sqrt{1-\eta^2} & -\eta & 0 \\ 0 & 0 & 1 \end{pmatrix}\,.
\end{align}
$\hat H_\text{ref}=\hbar\omega_\text{ref}\ket{1}\!\!\bra{1}$ is the Hamiltonian at the reference bias where $\omega_\text{ref}$ is the corresponding $0\leftrightarrow1$ transition frequency. Since we neglect higher excitation states in the shallow well, $P$ has one parameter $\eta$. Furthermore, $P$'s form results from unitarity and it induces Landau-Zener type physics between $\ket{0}$ and $\ket{1}$ \cite{Landau_PZS_2_46, Zener_PRC_137_696}. Indeed, if the pulse is non-adiabatic, i.e. it contains rapid changes in flux bias, state transitions can occur. They result from the non-orthogonality between the wave functions of the new excited state and old ground state; There exists a matrix element connecting the two states. On the other hand, if the change is slow, i.e. adiabatic, then the state cannot jump between eigenstates and remains in its initial state. The matrix element connecting ground and excited state is negligibly small at all points in time. This effect is modeled by choosing $\eta$ to be the overlap between the wave function $\psi_0$ of $\ket{0}$ at the reference bias and itself at a different bias
\begin{align}\notag
 \eta(\varphi_\text{b})=\int\psi_0^*(\varphi,\varphi_\text{b})\,\psi_0(\varphi,\varphi_\text{ref})\,{\rm d} \varphi\,.
\end{align}
The wave functions are found with a discrete variable representation (DVR) \cite{Colbert_JCP_96_19821991}. This consists of diagonalizing the phase qubit Hamiltonian (\ref{Eqn:PII_PhaseQubitH}) in a discretized eigenbasis of $\hat\varphi$ for different flux biases. The resulting eigenvalues are the energy levels and the associated eigenvectors are the wave-functions as function of phase $\varphi$. This yields $\eta$ which is then fitted to a third order polynomial, see Fig. \ref{Fig:PII_3lvl_eta}. The fit to a polynomial preserves the analytical aspect of the gradient computation.

\begin{figure} \centering
 \includegraphics[width=0.95\columnwidth]{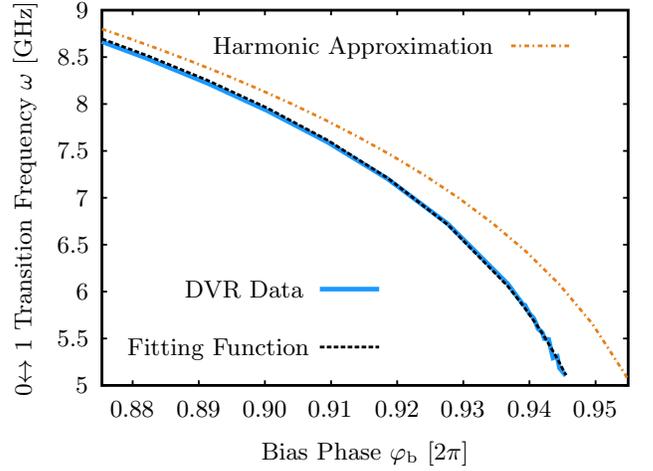}
 \caption{Frequency of the $\ket{0}\leftrightarrow\ket{1}$ transition for the Harmonic approximation and for the DVR data which takes all orders into account. The dashed line is the fit to the five parameter function $a(b+c\varphi_\text{b})^d+e$ showing excellent agreement for the range of bias flux of concern. Note that beyond $\varphi_\text{b}=0.945\cdot2\pi$ DVR no longer finds two states in the shallow well. This is in excellent agreement with the three level model validity condition shown in Fig. \ref{Fig:PII_3lvl_model}. \label{Fig:DVR_vs_Harmonic} }
\end{figure}

The Lindblad operators of the incoherent processes that we include in our model are
\begin{align}
 \hat L_{0\to\text{m}}=\sqrt{\gamma_0}\ket{\text{m}}\!\!\bra{0} \notag \\
 \hat L_{1\to\text{m}}=\sqrt{\gamma_1}\ket{\text{m}}\!\!\bra{1} \notag \\
 \hat L_{1\to0}=\sqrt{\gamma_{1\to0}}\ket{0}\!\!\bra{1} \notag
\end{align}
Note that we do not include pure dephasing between $\ket{0}$ and $\ket{1}$ since the coherences between these states do not matter when it comes to the measurement process. Whilst the relaxation rate $\gamma_{1\to0}=T_1^{-1}$ is constant, the tunneling rates to the continuum $\gamma_0$ and $\gamma_1$ depend on the bias flux. They are found by approximating the potential well by a third order polynomial \cite{Neeley_NatPhys_4_523} and using the WKB approximation \cite{WeissBook, Weiss_PRD_27_2916}
\begin{align}
 \gamma_0(\alpha)\simeq&~6\omega\sqrt{\frac{\alpha}{\pi}}e^{-\frac{6}{5}\alpha}\,, \notag \\
 \gamma_1(\alpha)\simeq&~432\omega \sqrt{\frac{\alpha^3}{\pi}}e^{-\frac{6}{5}\alpha}\,. \notag
\end{align}
$\omega$ is the $0\leftrightarrow1$ transition frequency in the harmonic approximation. This frequency is obtained from the second order term of the third order approximation employed by Martinis \emph{et al.} \cite{Martinis_PRL_55_1543} by building on work done by Caldeira and Leggett \cite{Caldeira_AP_149_374}. We improve this approximation for $\omega$ by using the DVR of the potential and then finding the eigenenergies for $\ket{0}$ and $\ket{1}$ in the shallow well at different bias values. The DVR data for $\omega$ is fitted to the five parameter function $a(b+c\varphi_\text{b})^d+e$ so that analytical gradients can be computed. This methodology, shown in Fig. \ref{Fig:DVR_vs_Harmonic}, allows for a good fit to the DVR data and shows that the harmonic approximation deviates a little from it. This is expected since by diagonalizing the Hamiltonian in a phase basis, DVR takes all orders of the potential into account.

\begin{figure} \centering
 \includegraphics[width=0.95\columnwidth]{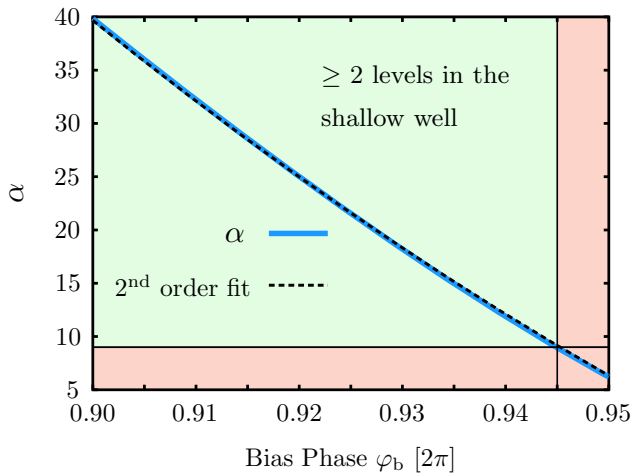}
 \caption{Parameter controlling the tunneling rates in the three level model. The solid line shows the value of $\alpha$ as computed by Eq. (\ref{Eqn:PII_PQ_alpha}) whilst the dashed line shows a second order fit used to preserve analyticity during the gradient search. When $\alpha$ goes below the horizontal line, the three level model is no longer valid. \label{Fig:PII_3lvl_model}}
\end{figure}

The dimensionless parameter $\alpha$ also depends on the bias flux. In the cubic potential model it is given by
\begin{align} \label{Eqn:PII_PQ_alpha}
 \alpha(\varphi_\text{b})=6\frac{V_\text{max}-V_\text{min}}{\sqrt{2E_JE_c}(\beta^{-1}+\cos\varphi_\text{min})}\,.
\end{align}
The potential extrema $V_\text{max/min}$ are defined in Fig. \ref{Fig:PII_PhaseQubitPot}. The phase value corresponding to the minimum is $\varphi_\text{min}$. These quantities all depend on the bias flux. The derivation of this formula is based on the expression of $\alpha$ found from the WKB approximation and the parameters entering the third order approximation of the qubit's potential. Some additional details are given in appendix \ref{Sec:Appendix_PQ}. Although not explicitly indicated, the potential extrema $V_\text{min/max}$ and the location of the minimum $\varphi_\text{min}$ depend on the bias flux. $\alpha$ is found numerically by solving for the different terms in Eq. (\ref{Eqn:PII_PQ_alpha}) for different values of $\varphi_\text{b}$. The result is shown in Fig. \ref{Fig:PII_3lvl_model} the numerical data is then fitted to a second order polynomial to preserve analyticity when computing gradients for the pulse optimization. In summary, the drift generator of the time propagator is
\begin{align} \notag
 \mathcal{S}_\text{d}=\gamma_{1\to0}\LR{\ket{0}\!\!\bra{1}\!\otimes\!\ket{0}\!\!\bra{1}-\frac{1}{2}\LR{\ket{1}\!\!\bra{1}\!\otimes\!\mathds{1}+\mathds{1}\!\otimes\!\ket{1}\!\!\bra{1}}}.
\end{align}
The control generator is
\begin{align} \notag
 &\mathcal{S_\text{c}}=i\LR{(P\hat H_\text{ref}P^{-1})^T\otimes\mathds{1}-\mathds{1}\otimes P\hat H_\text{ref}P^{-1} }+\\
 &\sum\limits_{j=0}^1\gamma_j(\varphi_\text{b})\!\LR{\ket{\text{m}}\!\!\bra{j}\!\otimes\!\ket{\text{m}}\!\!\bra{j}-\frac{1}{2}\LR{\ket{j}\!\!\bra{j}\!\otimes\!\mathds{1}+\mathds{1}\!\otimes\!\ket{j}\!\!\bra{j}}}. \notag
\end{align}
In the first term, the dependence on the bias flux is located in the $\eta$ parameter in the unitary matrix $P$. The non-linearity of this expression in the control $\varphi_\text{b}$ can easily be taken into account in the optimization using the chain rule.

\subsubsection{Optimal Control Problem}
The control problem is to optimize a measurement pulse $\varphi_\text{b}(t)$ of duration $T_\text{meas}$ that maximizes the contrast $\xi=P_\text{bright}(1-P_\text{dark})$ \cite{Chen_PRL_107_217401}. $P_\text{bright}$ is the probability that the initial state $\ket{1}$ tunneled to $\ket{m}$ whilst $P_\text{dark}$ is the probability that the state $\ket{0}$ tunneled to $\ket{m}$. This target can be shaped into a Choi matrix given by
\begin{align}\notag
 C_\text{t}=\ket{1}\!\!\bra{1}\otimes\ket{\text{m}}\!\!\bra{\text{m}}+\!\!\sum\limits_{i,j\in\{0,\text{m}\}}\!\!\ket{i}\!\!\bra{j}\otimes\ket{i}\!\!\bra{j}\,.
\end{align}
Since the tunneling is incoherent the coherences between $\ket{1}$ and $\ket{\text{m}}$ are not preserved as the ideal quantum channel maps $\ket{1}\!\!\bra{1}$ to $\ket{\text{m}}\!\!\bra{\text{m}}$. This gives the first part of $C_\text{t}$. The second states that the elements $\ket{0}\!\!\bra{0}$, $\ket{\text{m}}\!\!\bra{0}$, $\ket{0}\!\!\bra{\text{m}}$ and $\ket{\text{m}}\!\!\bra{\text{m}}$ should be left untouched.

Before and after the measurement pulse, the qubit is at a reference bias $\varphi_\text{ref}$ chosen such that tunneling out of $\ket{1}$ is suppressed. Indeed it is expected that coherent operations are done between $\ket{0}$ and $\ket{1}$ before the measurement pulse. Therefore the states should not tunnel out of the shallow well. However the shape of the wave-functions $\psi_i(\varphi,\varphi_\text{b})=\braket{\varphi|i}$ for $i=0,1$ change with bias flux. Thus changing $\varphi_\text{b}$ can induce $\ket{0}\leftrightarrow\ket{1}$ transitions if it is non-adiabatic, similar to the Landau-Zener scenario. This, as well as tunneling from $\ket{0}$ to $\ket{\text{m}}$,  creates dark counts. To avoid such effects an adiabatic pulse, with the appropriate area to minimize $\ket{0}\to\ket{\text{m}}$, should be used since slow changes in the potential will keep the system in $\ket{0}$ if it started in $\ket{0}$. However, $\ket{1}\to\ket{0}$ relaxation, graphically illustrated in Fig. \ref{Fig:PII_PhaseQubitPot}, causes missed counts. This degradation in contrast can be mitigated by using a fast pulse. This interplay between Landau-Zener like behavior and energy relaxation prompts the use of optimal control theory to shape the measurement pulse. The optimal pulse should reduce dark and missed counts. The former are reduced by the optimal shape whilst that latter are mitigated by forcing $\ket{1}$ to tunnel before relaxation happens.

\subsubsection{Baseline}
State measurement with phase qubits was originally limited by the high amount of two level fluctuators polluting the qubit \cite{Cooper_PRL_93_180401}. This has been overcome and phase qubit measurement visibilities around 90\% have been reported \cite{Steffen_PRL_97_050502, Chen_APL_101_182601}. Single photon measurement contrasts with JPMs approach 80\% \cite{Chen_PRL_107_217401}. Within the framework of the simplified model presented here the following section shows that these numbers could be increased. The limitations of a three level model could be overcome using Ad-HOC \cite{Egger_PRL_112_240503}, a closed-loop fine-tuning approach for pulses. The optimized pulses presented in the following section should thus be understood as a starting point for a closed-loop algorithm.

\subsection{Optimization Results}
\begin{figure} \centering
 \includegraphics[width=0.95\columnwidth]{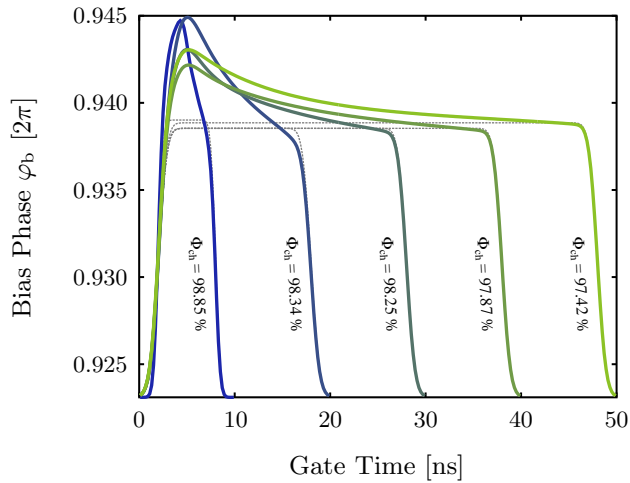}
 \caption{Optimal pulses for different gate durations with their corresponding fidelities. The dashed lines show the initial guess. As can be seen the fidelity of the optimal pulses is higher for the fast pulses. This is due to the fact that faster pulses allow $\ket{1}$ to tunnel into $\ket{\text{m}}$ before $T_1$ relaxes it to $\ket{0}$. \label{Fig:PII_Ch6_T1500Pulses}}
\end{figure}

\begin{figure} \centering
 \includegraphics[width=0.95\columnwidth]{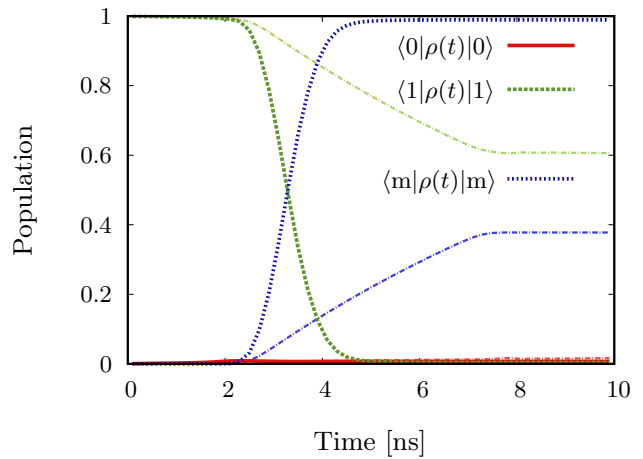}
 \caption{Time evolution of the populations for the 10 ns pulse of Fig. \ref{Fig:PII_Ch6_T1500Pulses} starting from the $\ket{1}\!\!\bra{1}$ state. As can be seen the initial pulse (corresponding to the thin lines) has a non-optimal pulse that fails to transfer population to $\ket{\text{m}}$ which would result in missed counts. Its channel fidelity and contrast are respectively $\Phi_\text{ch,i}'=87.0\%$ and $\xi_\text{i}=37.8\%$. The optimal pulse (thick lines) corrects  for this as well as preventing population transfers between $\ket{0}$ and $\ket{1}$. It has $\Phi_\text{ch,f}'=98.8\%$ and $\xi_\text{f}=97.9\%$. \label{Fig:PII_Ch6_T1500nsPopul} }
\end{figure}

The parameters used in the optimization correspond to typical phase qubit values \cite{Neeley_NatPhys_4_523}. These are shown in Tab. \ref{Tab:PII_PQVal}. Sharp jumps in the bias flux can introduce unwanted $\ket{0}\leftrightarrow\ket{1}$ jumps and cause St\"uckelberg oscillations, i.e. oscillations typical for finite-amplitude parameter sweeps \cite{Stueckelberg_HPA_5_369}. To prevent this, the pulses are convoluted with a Gaussian. This also results in pulses that are feasible with modern electronics. The optimization of several pulses of variable time is shown in Fig. \ref{Fig:PII_Ch6_T1500Pulses}. The initial guess for the gradient search is a square pulse convoluted by a Gaussian. The first and last two ns are held constant and only change through the convolution due to variations in the optimization pixels. The height of the initial pulse is too low to allow full tunneling out of $\ket{1}$ yet sufficiently high for small changes in the amplitude to produce appreciable changes in fidelity. This is best seen in Fig. \ref{Fig:PII_Ch6_T1500nsPopul} where the time evolution resulting from a 10 ns pulse is shown. The initial pulse fails to let the $\ket{1}$ state tunnel out. The initial channel fidelity and contrast are respectively $\Phi_\text{ch,i}'=87.0\%$ and $\xi_\text{i}=37.8\%$. After optimization these numbers are $\Phi_\text{ch,f}'=98.8\%$ and $\xi_\text{f}=97.9\%$ and show that the optimization has successfully increased the contrast, as desired.

\begin{table}
 \begin{center}
  \caption{Values used in the phase qubit model.\label{Tab:PII_PQVal}}
  \begin{tabular}{l l r l} \hline\hline
   Name & Symbol & Value & unit \\ \hline
   Critical Current & $I_0$ & 2 & $\mu A$ \\
   Junction Capacitance & $C$ & 1 & $pF$ \\
   Flux coupling & $\beta$ & 4.375 & - \\
   Energy Relaxation & $T_1$ & 500 & ns \\\hline\hline
  \end{tabular}
 \end{center}
\end{table}

The optimization adds a bump on the initial rise of the pulse to kick out the $\ket{1}$ state. This bump has to be added at the beginning of the pulse before $T_1$ relaxes $\ket{1}$ to $\ket{0}$ which should be kept in the shallow well. The optimization carefully choses the area under the pulse. Indeed, allowing the bias flux to be held too high for too long diminishes the contrast since $\ket{0}$ starts to tunnel into $\ket{\text{m}}$. The rate at peak is pushed close to the maximum allowed by the three level model. This limitation is further discussed below. The slow hold value at the end of the longer pulses in Fig. \ref{Fig:PII_Ch6_T1500Pulses} is an artifact of the simulation resulting from the initial guess. For shorter pulses no hold value subsists. In the longer pulses it remains since it does not affect fidelity. Indeed, after the kick forcing $\ket{1}$ to tunnel, there can be no fidelity deterioration due to $T_1$. Furthermore this artificial hold level is too low to allow any significan tunneling of $\ket{0}$ into $\ket{\text{m}}$.

\begin{figure} \centering
 \includegraphics[width=0.98\columnwidth]{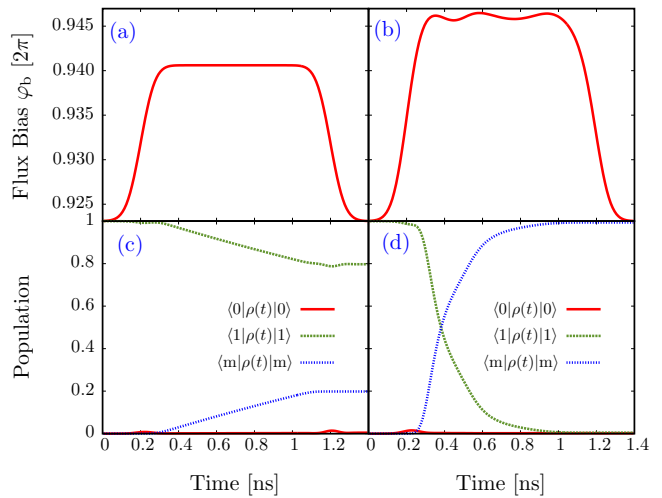}
 \caption{Optimization of a fast readout pulse. \textcolor{blue}{(a)} initial pulse sequence with fidelity $\Phi_\text{ch,i}'=83.8\%$ and contrast $\xi_\text{i}=19.8\%$. \textcolor{blue}{(b)} Optimized pulse shape. \textcolor{blue}{(c)} Initial time evolution of populations. Again, the unoptimized pulse fails to let $\ket{1}$ tunnel into $\ket{m}$. \textcolor{blue}{(d)} Time evolution of populations after pulse optimization resulting in a high contrast of $\xi_\text{f}=98.2\%$ and final fidelity $\Phi_\text{ch,f}'=99.2\%$. \label{Fig:PII_Ch6_TFastPulses}}
\end{figure}

Faster pulses than those in Fig. \ref{Fig:PII_Ch6_T1500Pulses} were optimized. A 1.4 ns pulse is shown in Fig. \ref{Fig:PII_Ch6_TFastPulses} the initial fidelity and contrast were $\Phi_\text{ch,i}'=87.0\%$ and $\xi_\text{i}=37.8\%$, whilst the optimized pulse has $\Phi_\text{ch,f}'=98.8\%$ and $\xi_\text{f}=97.9\%$. However, faster pulses cannot be made in this model since it relies upon having at least two states in the meta stable well. This imposes a restriction on the maximum bias flux. Approximating the potential with a third order polynomial and asking for at least two levels in the well leads to the approximate condition $\alpha>9$ (details are in appendix \ref{Sec:Appendix_PQ}). This threshold value is shown by the horizontal line in Fig. \ref{Fig:PII_3lvl_model} and corresponds to a flux bias of $0.9454\cdot2\pi$. Also note that this value matches very well the maximum bias for which DVR can still find at least two states in the shallow well, see Fig. \ref{Fig:DVR_vs_Harmonic}. In the pulse optimization, the flux bias is constrained to be below this value. Thus, upon examining the optimal pulse in Fig. \ref{Fig:PII_Ch6_TFastPulses} it can be seen that the pulse has reached this limit. Therefore the tunneling rate out of $\ket{1}$ has reached its maximum within the validity of the three level model. It may thus be possible to extend contrast even further by biasing so that the excited state falls into the continuum. However theoretical description of this regime falls way beyond the scope of this paper.

\section{Outlook and Conclusions \label{sec:conclusion}}
Optimal control in the presence of non-unitary dynamics towards a target unitary time evolution has already been implemented. In this work we have taken this a step further and presented a methodology to optimize a non-unitary time evolution towards a non-unitary target channel using a gradient search on a fidelity measure based on the Choi matrix. The algorithm was illustrated within the framework of optimizing a measurement pulse for a phase qubit where the measurement process relies on incoherent tunneling processes. The simple model shows a rich interplay between Landau-Zener type physics and the incoherent dynamics. The three level model discussed here is a good starting point for creating a measurement pulse. Going beyond this model could be done in the experiments by using the methodology developed in \cite{Egger_PRL_112_240503}. Measurement is important for superconducting qubits. Optimizing pulses for different systems, such as dispersive readout through a resonator, will require additional developments in OCT and could be the topic of future research.

\section{Acknowledgments}
We thank John M. Martinis for useful discussions and Luke C. G. Govia for his careful reading of the manuscript. This work was supported by the Army Research Office under contract W911NF-14-1-0080 and the European Union through ScaleQIT. This research was also funded by the Office of the Director of National Intelligence
(ODNI), Intelligence Advanced Research Projects Activity (IARPA), through the Army Research Office. All statements of fact, opinion, or conclusions contained herein are those of the authors and should not be construed as representing the official views or policies of IARPA, the ODNI, or the US government.

\appendix

\section{Phase Qubit Potential \label{Sec:Appendix_PQ}}
The Hamiltonian of the flux biased phase qubit is
\begin{align} \notag
 \hat H=E_c\hat N^2+\underbrace{E_J\LR{\frac{1}{2\beta}(\hat \varphi-\varphi_\text{b})^2-\cos\hat\varphi}}_{V_J~\text{qubit potential}}\,.
\end{align}
This Hamiltonian can be approximated by a third order potential $\hat V_3$ with three parameters $m$, $\omega$ and $\tilde\varphi$. The approximate Hamiltonian is given by
\begin{align} \notag
 \hat H'=\frac{\hbar^2}{2m}\hat N^2+\frac{1}{2}m\omega^2\Delta\varphi^2\LR{1-\frac{2}{3}\frac{\Delta\varphi}{\tilde\varphi}}\,.
\end{align}
The phase variable is $\Delta\varphi=\varphi-\varphi_\text{min}$ where $\varphi_\text{min}$ is the local minimum of the shallow well. By comparing $\hat H$ and $\hat H'$ it is straight forward to identify the effective mass as $m=\hbar^2/2E_c$. Note that this parameter has units of energy instead of mass. In this approximation the constant term in the potential has been dropped so that $V_3(\Delta\varphi=0)=0$. The frequency $\omega$ in the third order potential is chosen such that the harmonic term matches the second derivative of the actual phase qubit potential
\begin{align} \label{Eqn:A1_1}
 \frac{1}{2}V_J''(\varphi_\text{min})=&~\frac{1}{2}m\omega^2 \\ \Longrightarrow~\hbar\omega=&~\sqrt{2E_cE_J(\beta^{-1}+\cos\varphi_\text{min})}\,.\notag
\end{align}
Lastly, $\tilde\varphi$ is determined so that the potential barrier has the right hight. This imposes $V_3(\Delta\varphi=\tilde\varphi)=V_\text{max}-V_\text{min}$ where $V_\text{min/max}$ are the local minima/maxima close to the shallow well, see Fig. \ref{Fig:PII_PhaseQubitPot}. This leads to the following expression
\begin{align} \label{Eqn:A1_2}
 m\omega^2\tilde\varphi=6(V_\text{max}-V_\text{min})\,.
\end{align}
The tunneling rates given in chapter 12 of Weiss \cite{WeissBook} for the potential $V_3$ involve the term $m\omega^2\tilde\varphi^2/\hbar\omega$ which, for brevity, was labeled $\alpha$ in the main text. Therefore, combining Eqs. (\ref{Eqn:A1_1}) and (\ref{Eqn:A1_2}) yields
\begin{align} \notag
 \alpha=6\frac{V_\text{max}-V_\text{min}}{\sqrt{2E_JE_c}(\beta^{-1}+\cos\varphi_\text{min})}\,.
\end{align}
The tree level model is valid if the shallow well contains at least two states. This imposes that the barrier height $V_3(\tilde\varphi)$ be greater than $3\hbar\omega/2$. Using the definition of $\alpha$ results in the condition $\alpha>9$ for the three level model to be valid. 

\bibliographystyle{apsrev4-1}
\bibliography{QC_PublicationDataBase}

\end{document}